\documentclass[showpacs,preprintnumbers,amsmath,amssymb,nofootinbib,twocolumn]{revtex4-1}

\usepackage[small]{caption}
\usepackage{amsmath,amsfonts,amscd,amssymb,epsf, epsfig,mathrsfs,color}\usepackage{graphicx}

\newcommand{\pa}{\partial}
\newcommand{\vep}{\varepsilon}

\begin{document}
\title{Bubble-wall Casimir interaction in fermionic environments}
\author{Antonino Flachi}
\affiliation{Centro Multidisciplinar de Astrof\,$\acute{\imath}$sica, Departamento de Fis\'ica, Instituto Superior T$\acute{\text{e}}$cnico, Universidade de Lisboa, Avenida Rovisco Pais 1, 1049-001 Lisboa, Portugal.}
\author{Lee-Peng Teo}
 \affiliation{Department of Applied Mathematics, Faculty of Engineering, University of Nottingham Malaysia Campus, Jalan Broga, 43500, Semenyih, Selangor Darul Ehsan, Malaysia.}

\begin{abstract}
We consider the Casimir interaction, mediated by massless fermions, between a spherical defect and a flat potential barrier, assuming hard (bag-type) boundary conditions at both the barrier and the surface of the sphere. The computation of the quantum interaction energy is carried out using the multiple scattering approach, adapted here to the setup in question.
We find an exact integral formula for the energy, from which we extract both the large and short distance asymptotic behaviour. At large distance the fermionic contribution is found to scale as $L^{-3}$, in contrast to that of electromagnetic vacuum fluctuations that, assuming perfectly conducting boundaries, scales as $L^{-4}$. At short distance, we compute the leading and sub-leading contribution to the vacuum energy. The leading one coincides with what it is expected from the proximity force approximation, while the sub-leading term gives, contrary to the electromagnetic case, a positive correction to the proximity force result.
\end{abstract}
 \maketitle

\emph{Introduction.} In 1948 Casimir predicted that two flat, and perfectly conducting plates would modify the vacuum fluctuations of the electromagnetic field and induce an observable force between them \cite{casimir}. This set-up was difficult to unambiguously test experimentally at the time \cite{sparnay}, and the first conclusive observation came many years later, in fact, for the different configuration of a sphere near a plate \cite{lamoreaux}. It took some more time to conclusively observe the Casimir force between two plates, which was, eventually, achieved a few years later \cite{bressi} (See Ref. \cite{review} for review).

Much of the work that followed Casimir's conclusion focused on the electromagnetic vacuum fluctuations and different geometries (See Refs. \cite{review}). However, the parallel plates setup was just one specific example of the more generic deformations in the quantum vacuum that boundaries can produce, irrespectively of the nature of the quantum fields. The key feature is the presence of massless (or quasi-massless) quanta that induce long-ranged correlations, suggesting that quantum vacuum effects may be relevant in fermionic environments. The analogous of the electromagnetic Casimir effect should, in fact, occur in condensed matter systems, like in quantum liquids \cite{volovik}, or when long-range correlations exist due to Goldstone modes of a broken continuous symmetry, as in superfluids \cite{kardar,schecter}.

One particularly exciting example is associated with the presence of defects in quantum fermi liquids, and with the possibility that high precision experimental manipulation of ultra-cold atomic systems \cite{bloch}, where defects can be controlled in a variety of ways, may offer novel tests of quantum vacuum energy effects and, for instance, provide new constraints on hypothetical sub-micron interactions. Other physical setups relevant to the fermion Casimir effect include, for instance, carbon nanotubes \cite{zhabinskaya,Elizalde:2011cy}, graphene \cite{fialk,flachi}, nuclear structures \cite{nuclear}, neutron star crusts \cite{Bulgac:2000qk} (See Refs.~\cite{recati} for a longer list of examples and references).

The simplest setups analysed have, so far, ignored the structure of the defects and focused on one-dimensional boson and fermi systems with defects treated as delta-functions in the adiabatic approximation (See, for example, Refs.~\cite{recati,zhabinskaya,Sundberg:2003tc}). In Refs.~\cite{recati}, it was found that, for a generic (interacting or noninteracting) fermionic background, the Casimir force between the impurities oscillates as a function of the separation. The similar problem of calculating the Casimir interaction between two scatterers immersed in a one-dimensional massless fermionic background has also been analysed in Ref.~\cite{zhabinskaya}, using a force operator approach, with the defects modelled by two delta-function potentials. A similar oscillatory behaviour has been found too, however, as a function of relative polarization of the two scatterers, while the dependence on the distance turned out to be monotonic (for fixed polarizations).

A fuller understanding of the Casimir effect mediated by fermions should include the structure of the defect and extend to higher dimensionality.
This class of problems has been analysed, for instance, in Refs.~\cite{bulgac}, where the properties of systems containing one or more fermionic bubbles (almost spherical defects immersed in a homogenous fermionic environment) have been discussed. The Casimir energy of a system composed of two, three, and four spheres in a fermionic background has been analysed in Ref.~\cite{bulgac} in the semiclassical approximation and led to the suggestion that in many-body Casimir interactions, the two-body ones dominate at small separation.

In the present work, we wish to consider the related problem of computing the Casimir interaction between a defect
and a potential barrier, encapsulating a massless fermionic system. Following Ref.~\cite{bulgac}, we model the defect as a spherical bubble, and both the bubble and the barrier as hard walls. The geometry of the system is that of a sphere of radius $R$, centered at the origin, close to a plate of surface area $H\times H$ located at $z=L$, at a distance $d$ from the sphere. Since we are considering the case of a defect whose size is much smaller than the surface area of the wall, we shall assume that $R\ll H$.

Despite the simplicity of the setup, the problem of studying the Casimir interaction between a sphere and a wall  is not straightforward and has been subject of many analyses that have focused on the scalar and electromagnetic quantum vacuum fluctuations \cite{n1,5,6,2,3,4,7,8,9}. While the leading term in this interaction at short distance can be obtained easily using the proximity force approximation \cite{derjaguin}, going beyond is difficult.

The increasing precision at which Casimir force measurements can be performed called for more accurate computations and for efficient ways to go beyond the proximity force approximation. An especially advantageous one is based on the multiple scattering approach and has offered a systematic way to compute the Casimir energy between two compact objects of arbitrary shape \cite{n1,5,6,2,3,4,7,8,9}. Results have covered a variety of cases, mainly for the scalar and electromagnetic fields (See, for instance, Refs.~\cite{2,3,4,7,8,mara} or Ref.~\cite{9} for review). The beauty of the multiple scattering approach is that it is physically transparent, since the interaction energy is expressed in terms of a multiple scattering expansion (waves that scatter back and forth between the two objects that are interacting). It is also of straightforward numerical implementation, and it allows for a systematic way to extract sub-leading corrections to proximity force results.

For the sphere-plate configurations, the multiple scattering approach has been adopted to obtain the scalar Casimir energy beyond the proximity force approximation has been worked out in Refs.~\cite{12,14}. Results can be summarised in the formula ${E^{\text{XY}}/E_{\text{PFA}}^{\text{XY}}} \approx 1 + \sigma^{\text{XY}} d/R$, where X and Y represent the boundary conditions imposed at the plate (X = Dirichlet (D), Neumann (N), Robin (R)) and at the sphere (Y = D, N, R) , respectively, and $E_{\text{PFA}}^{\text{XY}} = \rho^{\text{XY}} R/d^2$. The numerical coefficients are $\rho^{\text{DD}/ \text{NR}} = -\pi^3/1440$, $\rho^{\text{ND}/ \text{DR}} = -7 \pi^3/11520$,  $\sigma^{\text{DD}/ \text{ND}} = 1/3$, $\sigma^{\text{DR}}=1/3+80(3a-2)/(7\pi^2)$ and $ \sigma^{\text{NR}} =1/3+ 20(3 a - 2)/\pi^2$ ($a$ is the Robin   parameter). For the electromagnetic field, the Casimir energy beyond the proximity force approximation has been computed in Refs.~\cite{13,14} for perfectly conducting boundary conditions, leading to $E^{\text{EM}} \approx - \pi^3 R/(720d^2)(1+(1/3-20/\pi^2)d/R)$.

\emph{Fermion TGTG formula.} While the multiple scattering approach has been systematised for the electromagnetic and scalar Casimir effect, only limited attention has been paid to the fermionic case.  The possibility of high precision Casimir effect experiments in fermionic environments (for instance, in ultra-cold atomic systems \cite{bloch}), provides a natural motivation to carry out more precise computations, beyond proximity force results.

As shown in Refs.~\cite{6,2,3,4}, the multiple scattering approach allows to express the Casimir energy in terms of transition matrices (associated to the interacting bodies) and the propagators, thus offering a precise computational prescription, and, in the following, we adapt this approach to our case. The formal expression for the fermion Casimir energy takes the usual form
\begin{equation}\label{eq11_18_2}
E_{\text{Cas}}=-\frac{\hbar  }{2\pi}\int_0^{\infty} d\xi \ln\det\left(\mathbb{I}
-
\mathbb{N}
\right),
\end{equation}
where $\mathbb{N} \equiv \mathbb{T}^1\mathbb{G}^{12}\mathbb{T}^2\mathbb{G}^{21}$ and $\xi$ is the imaginary frequency. The matrices $\mathbb{T}^i$, $i=1,~2$ are the transition matrices associated to the boundaries (the sphere and the wall in our case) and $\mathbb{G}^{ij}$ represent the translation matrices. Formally, aside for the change in the overall sign, in the expression above (also called TGTG-formula after the work \cite{6}), the nature of the quantum fields is encoded in the matrices inside the determinant. The matrices $\mathbb{T}^i$ are related to the scattering matrix of object $i$ and can be computed by matching the boundary conditions imposed on object $i$. The translation matrices $\mathbb{G}^{ij}$ relate the basis of wave functions of object $i$ to the basis of wave functions of object $j$. Expression (\ref{eq11_18_2}) is valid at a formal level and deriving an explicit expression reduces to calculating the above matrices and taking determinant. Here we will follow the procedure outlined in Ref.~\cite{9}, where a prescription to compute the matrices $\mathbb{T}^i$ and $\mathbb{G}^{ij}$ using the mode-summation approach has been developed.

The fermions in our computation are massless spin $1/2$-fermionic fields $\psi$, satisfying the Dirac equation, $i\gamma^{\mu}\nabla_{\mu}\psi=0$, where
$\nabla_{\mu}=\pa_{\mu}+\Gamma_{\mu}$, and $\Gamma_{\mu}$ is the spin connection. On the boundaries of the sphere ($B=$ sphere) and the wall ($B=$ wall), we impose $(1+i\gamma^{\mu}n_{\mu})\psi\Bigr|_{B}=0$, with $n_{\mu}$ being the unit outward normal vector.

In order to match the boundary conditions and find the transition matrices, one needs to express the solutions in terms of a spherical and plane wave basis, respectively for the sphere and the plate. Explicit forms are known for spherical waves (See Ref.~\cite{10}):
\begin{equation}\label{eq5_16_3}
\begin{split}
\psi_{jm,1}^{(\pm), *}=&\mathcal{C}_j^{*} e^{\mp i\omega t}  \begin{pmatrix}f^*_{j-\frac{1}{2}}(kr)\Omega_{j,j-\frac{1}{2},m}\\
\mp i f^*_{j+\frac{1}{2}}(kr)\Omega_{j,j+\frac{1}{2},m}\end{pmatrix},\\
\psi_{jm,2}^{(\pm), *}=&\mathcal{C}_j^{*}e^{\mp i\omega t} \begin{pmatrix}f^*_{j+\frac{1}{2}}(kr)\Omega_{j,j+\frac{1}{2},m}\\
\pm i f^*_{j-\frac{1}{2}}(kr)\Omega_{j,j-\frac{1}{2},m}\end{pmatrix},
\end{split}
\end{equation}
where each mode is characterized by the quantum numbers $(j,m)$, with $j=1/2, 3/2, 5/2, \ldots$ and $m=-j, -j+1, \ldots, j-1, j$. The superscripts ${(+)}$ and ${(-)}$ indicate, respectively, the positive and negative energy modes, $\Omega_{jlm}$ represent the spherical harmonic spinors (see Ref.~\cite{10}), and $k=\omega/c$. For convenience, we have adopted the notation $*=$ reg ($*=$ out) for regular (outgoing) waves,
\begin{equation}\label{eq5_16_4}
f^{\text{reg}}_l(z)=\sqrt{\frac{\pi}{2z}}J_{l+\frac{1}{2}}(z),\quad f^{\text{out}}_l(z)=\sqrt{\frac{\pi}{2z}}H^{(1)}_{l+\frac{1}{2}}(z),
\end{equation}
with $\mathcal{C}_j^{\text{reg}}=i^{-j+\frac{1}{2}},\quad \mathcal{C}_j^{\text{out}}=\pi i^{j+\frac{3}{2}}/2$.
The plane waves can be parametrized in terms of the momenta perpendicular to the plate,  $\mathbf{k}_{\perp}=(k_1,k_2)$,
\begin{equation}\label{eq5_16_5}
\psi_{\mathbf{k}_{\perp},\alpha}^{(\pm), *}= A_{\mathbf{k}_{\perp},\alpha}^{(\pm), *} e^{ik_1x+ik_2y-i \text{sgn}_*\sqrt{k^2-k_{\perp}^2}z\mp i\omega t},
\end{equation}
where
$k_{\perp}=\sqrt{k_1^2+k_2^2}$, $\text{sgn}_{\text{reg}}=1$,  $\text{sgn}_{\text{out}}=-1$, and
\begin{equation}\label{eq5_16_6}\begin{split}
A_{\mathbf{k}_{\perp}, 1}^{(\pm), *}=&\begin{pmatrix}1\\ 0\\ \displaystyle  \mp \text{sgn}_* \frac{\sqrt{k^2-k_{\perp}^2}}{k}\\ \pm \displaystyle\frac{k_1+ik_2}{k}\end{pmatrix}, \\
A_{\mathbf{k}_{\perp}, 2}^{(\pm), *}=&\begin{pmatrix}0\\1\\ \displaystyle\pm\frac{k_1-ik_2}{k}\\  \displaystyle\pm\text{sgn}_{*} \frac{\sqrt{k^2-k_{\perp}^2}}{k}\end{pmatrix}.
\end{split}\end{equation}
The matching procedure is tedious but straightforward. It consists in expressing one set of waves in terms of the other, with the coefficients of this transformation defining the transition matrices. Imposing the boundary conditions, and solving the resulting equation for the transitions coefficients allows one to find $\mathbb{T}^1$ and $\mathbb{T}^2$ (Details for this and all the other calculations are reported in appendix). These are block-diagonal matrices in $(j,m)$ and $\mathbf{k}_{\perp}$ respectively. The $(j,m)$-block of $\mathbb{T}^1$ is  a diagonal $2\times 2$ matrix of the form
\begin{equation*}
\mathbb{T}_{jm}^{(\pm)}=
\begin{pmatrix}
T_{jm}^{(\pm)} & 0\\ 0 &\overline{T_{jm}^{(\pm)} }
\end{pmatrix},
\end{equation*}
where
\begin{equation}\label{eq5_16_1}
T_{jm}^{(\pm)} =\frac{I_j(\kappa R)\mp i I_{j+1}(\kappa R)}{K_j(\kappa R)\pm i K_{j+1}(\kappa R)},
\end{equation}
with $\kappa=ik$. The $\mathbf{k}_{\perp}$-block of $\mathbb{T}^2$ is
\begin{equation}\label{eq5_16_2}
\mathbb{T}_{\mathbf{k}_{\perp}}^{(\pm)} =\mp i
\begin{pmatrix}
\cos\theta_k & \sin\theta_k e^{-i\varphi_k}\\-\sin\theta_ke^{i\varphi_k} & \cos\theta_k
\end{pmatrix},
\end{equation}
where $\theta_k$ and $\varphi_k$ are defined so that $ \sqrt{k^2-k_{\perp}^2}=k\cos\theta_k$, $k_1=k_{\perp}\cos\varphi_k$ and $k_2=k_{\perp}\sin\varphi_k$.

The translation matrices $\mathbb{G}^{12}$ and $\mathbb{G}^{12}$ are defined by the relations
\begin{equation}\begin{split}
\begin{pmatrix}\psi^{(\pm), \text{reg}}_{\mathbf{k}_{\perp}, 1 }(\mathbf{x}',\omega)\\\psi^{(\pm), \text{reg}}_{\mathbf{k}_{\perp}, 2 }(\mathbf{x}',\omega)\end{pmatrix} =&\sum_{j }\sum_{m } \mathbb{G}^{12, (\pm)}_{jm, \mathbf{k}_{\perp}}\begin{pmatrix}\psi_{jm,1}^{(\pm),\text{reg}}(\mathbf{x},\omega)\\ \psi_{jm,2}^{(\pm),\text{reg}}(\mathbf{x},\omega)\end{pmatrix},\\
\begin{pmatrix} \psi^{(\pm), \text{out}}_{jm, 1 }(\mathbf{x},\omega)\\\psi^{(\pm), \text{out}}_{jm, 2 }(\mathbf{x},\omega)\end{pmatrix} =&H^2
\int \frac{d^2k}{(2\pi)^2}
\mathbb{G}^{21, (\pm)}_{ \mathbf{k}_{\perp}, jm}
\begin{pmatrix}\psi_{\mathbf{k}_{\perp},1}^{(\pm),\text{out}}(\mathbf{x}',\omega)\\\psi_{\mathbf{k}_{\perp},2}^{(\pm),\text{out}}(\mathbf{x}',\omega)\end{pmatrix}
\end{split}\end{equation}
where $\mathbf{x}'=\mathbf{x}-\mathbf{L}$, $\mathbf{L}=(0,0,L)$. The computation can be performed following the idea introduced in \cite{11} for the computation of the corresponding translation matrices for scalar and electromagnetic fields. We find the following result :
\begin{widetext}
\begin{equation}\label{eq5_16_7}
\begin{split}
\mathbb{G}^{12, (\pm)}_{jm, \mathbf{k}_{\perp}}
=
&
(-1)^{-m+\frac{1}{2}}
\sqrt{4\pi}\sqrt{\frac{(j-m)!}{(j+m)!}} e^{-i\left(m-\frac{1}{2}\right)\varphi_k}e^{i\sqrt{k^2-k_{\perp}^2}L} \begin{pmatrix}
(j+m)P_{j-\frac{1}{2}}^{m-\frac{1}{2}}(\cos\theta_k)&-P_{j-\frac{1}{2}}^{m+\frac{1}{2}}(\cos\theta_k)e^{-i\varphi_k}\\i(j-m+1)P_{j+\frac{1}{2}}^{m-\frac{1}{2}}(\cos\theta_k)&
iP_{j+\frac{1}{2}}^{m+\frac{1}{2}}(\cos\theta_k)e^{-i\varphi_k}
\end{pmatrix},\end{split}
\end{equation}\begin{equation}\label{eq5_16_9}
\begin{split}
\mathbb{G}^{21, (\pm)}_{ \mathbf{k}_{\perp}, jm}
=&-\frac{\pi^{\frac{3}{2}}}{2H^2}\sqrt{\frac{(j-m)!}{(j+m)!}}\frac{e^{i\sqrt{k^2-k_{\perp}^2}L}}{k\sqrt{k^2-k_{\perp}^2}}e^{i\left(m-\frac{1}{2}\right)\varphi_k} \begin{pmatrix}(j+m)P_{j-\frac{1}{2}}^{m-\frac{1}{2}}(\cos\theta_k)&i(j-m+1)P_{j+\frac{1}{2}}^{m-\frac{1}{2}}(\cos\theta_k)\\P_{j-\frac{1}{2}}^{m+\frac{1}{2}}(\cos\theta_k)e^{i\varphi_k}&
-iP_{j+\frac{1}{2}}^{m+\frac{1}{2}}(\cos\theta_k)e^{i\varphi_k}
\end{pmatrix}.
\end{split}
\end{equation}
Combining the above expressions (using some identities for the associated Legendre functions $P_l^m(z)$, noticing that
the argument in the determinant is real, and the contribution from the positive energy modes and negative energy modes are the same), one can obtain the Casimir energy written as
\begin{equation}\label{eq5_16_8}
E_{\text{Cas}}=-\frac{\hbar  }{\pi}\int_0^{\infty} d\xi \ln\det\left(\mathbb{I} - \mathbb{M}\right),
\end{equation}
where
\begin{equation}\label{eq11_18_4}\begin{split}
M_{jm, j'm'}^{}
=&-\delta_{m,m'}
{i\pi \over 2}
\sqrt{\frac{(j-m)!(j'-m)!}{(j+m)!(j'+m)!}}\begin{pmatrix}
T_{jm}^{(+)} & 0\\ 0 &-\overline{T_{jm}^{(+)} }
\end{pmatrix}
\int_{0}^{\infty}d\theta\sinh\theta e^{-2\kappa L\cosh\theta}\times\\&\times \begin{pmatrix}(j+m)P_{j-\frac{1}{2}}^{m-\frac{1}{2}}(\cosh\theta)&P_{j-\frac{1}{2}}^{m+\frac{1}{2}}(\cosh\theta)\\ (j-m+1)P_{j+\frac{1}{2}}^{m-\frac{1}{2}}(\cosh\theta)&
-P_{j+\frac{1}{2}}^{m+\frac{1}{2}}(\cosh\theta)
\end{pmatrix}
\begin{pmatrix}(j'-m+1)P_{j'+\frac{1}{2}}^{m-\frac{1}{2}}(\cosh\theta)&(j'+m)P_{j'-\frac{1}{2}}^{m-\frac{1}{2}}(\cosh\theta)\\ P_{j'+\frac{1}{2}}^{m+\frac{1}{2}}(\cosh\theta) &-
P_{j'-\frac{1}{2}}^{m+\frac{1}{2}}(\cosh\theta)
\end{pmatrix},
\end{split}
\end{equation}
\end{widetext}
leading to an exact (integral) formula for the fermion Casimir energy between the wall and the sphere.

\emph{Asymptotic behaviour.} While (\ref{eq11_18_2})-(\ref{eq11_18_4}) can be used to compute numerically the fermion Casimir energy in our setup, here we are interested in the asymptotic behaviour at large, and, in particular, at short distance, both of which can be extracted from the main integral formula at the price of some lengthy computations.

First of all, we can compute the the Casimir interaction energy  at large separation, i.e., $L\gg R$. In this regime, the dominant contributions are those with $j=1/2$ and $m=\pm 1/2$, and a straightforward computation gives
\begin{equation}
E_{\text{Cas}} \approx -\frac{\hbar c R^2}{\pi L^3},
\end{equation}
leading to an attractive interaction at large distances. It seems interesting to notice that the fermion Casimir energy, at large distance, falls off as $L^{-3}$, less rapidly than the electromagnetic contribution that decays as $L^{-4}$. Compared to the scalar contribution, instead, the behaviour in the fermionic case is intermediate between that of a scalar with Dirichlet boundary conditions and that of a scalar with Neumann boundary conditions, in which cases the Casimir energy decays as $L^{-2}$ and $L^{-4}$, respectively.

The more interesting (and computationally more tedious) limit is that of the Casimir energy at small separation, $d\ll R$, beyond the leading (proximity force) approximation. In the present case, taking $j$ as the main quantum number and using the invariance of the matrix $M_{jm, j'm}$ under the change $m\mapsto -m$ (this implies that the next-to-leading order term is of order $d$ smaller than the leading term), a lengthy computation returns
\begin{equation}\label{eq11_18_6}\begin{split}
E_{\text{Cas}} = &-\frac{7\pi^3\hbar c R}{2880  d^2}\left(1+\left[\frac{1}{3}-\frac{20}{7\pi^2}\right]\frac{d}{R}+\ldots\right).
\end{split}\end{equation}
It is easy to check that the leading term above coincides with proximity force result, and gives rise to an attractive interaction, while the sub-leading term corrects the proximity force result by a positive amount. Contrary to what happens at large distance, in this limit the fermion and electromagnetic (for perfectly conducting boundaries) contributions are both attractive and scale in the same way. Interestingly, the correction to the proximity force result is, for the present setup, positive, in contrast to the analogous correction for the electromagnetic case with perfectly conducting boundaries, for which the correction is negative.

\emph{Conclusion}. The possibility of manipulating defects in condensed matter fermionic systems has triggered new curiosity in understanding the analogous of the Casimir energy in a fermionic environment and motivated the present work. In this paper, we have adapted the multiple scattering formalism to derive the Casimir interaction energy between a spherical defect and a wall mediated by massless fermionic quanta. We have obtained an integral representation for the quantum vacuum energy that is divergence free and valid at all distances. From this integral formula, we have extracted the leading contributions at both large and short distance. The behaviour of the Casimir energy at large distances scales as $L^{-3}$ and dominates over the electromagnetic contribution (for perfectly conducting boundaries), for which the energy scales as $L^{-4}$. The more interesting result comes from the short distance asymptotic behaviour, where the leading order contribution is found to coincide with the result obtained from the proximity force approximation. We have also derived the correction to the proximity force approximation, that, in contrast to the electromagnetic case, turns out to be positive. As a byproduct, we have derived the translation matrices, relating the plane waves basis to the spherical wave basis, a result that might be useful in other contexts. While the force is attractive at both small and large distance for the present choice of boundary conditions, it is important to ask how the result changes for different boundary conditions (for instance, introducing a phase at one of the boundaries), as well as when thermal effects are switched on. Work in this direction is in progress.

\acknowledgments

We acknowledge the support of   the Funda\c{c}\~{a}o para a Ci\^{e}ncia e a Tecnologia of Portugal and of the Marie Curie Action COFUND of the European Union Seventh Framework Program Grant Agreement No. PCOFUND-GA-2009-246542 (A.F.), and of the Ministry of Higher Education of Malaysia  under   FRGS grant FRGS/1/2013/ST02/UNIM/02/2 (L.-P.T.).

\begin{widetext}
\appendix

\section{I. Expansions}
A generic solution to the Dirac equation $\psi^{(\pm)}$ can be expressed as a superposition of the solutions presented in the main text (formulae (2) - (5)).
In the region between the sphere and the plane, $\psi^{(\pm)}$ can be represented in two ways. In terms of the full set of spherical solutions system (in spherical coordinates centered at $O$):
\begin{equation*}\begin{split}
\psi^{(\pm)}(\mathbf{x},t) =&\int_{-\infty}^{\infty} d\omega \sum_{j=\frac{1}{2},\frac{3}{2},\ldots}\sum_{m=-j, -j+1, \ldots, j-1, j}\\&\hspace{3cm}\times\left(a_{jm, 1}^{(\pm)} \psi_{jm,1}^{(\pm),\text{reg}}(\mathbf{x},\omega)+
a_{jm, 2}^{(\pm)} \psi_{jm,2}^{(\pm),\text{reg}}(\mathbf{x},\omega)+b_{jm, 1}^{(\pm)} \psi_{jm,1}^{(\pm),\text{out}}(\mathbf{x},\omega)+
b_{jm, 2}^{(\pm)} \psi_{jm,2}^{(\pm),\text{out}}(\mathbf{x},\omega)\right),\end{split}
\end{equation*}
or in terms of the full set of plane waves (in rectangular coordinates centered at $O'=L\mathbf{e}_z$):
\begin{equation*}\begin{split}
\psi^{(\pm)}(\mathbf{x}',t) =&H^2\int_{-\infty}^{\infty} d\omega \int_{-\infty}^{\infty}\frac{dk_x}{2\pi}\int_{-\infty}^{\infty}\frac{dk_y}{2\pi}\\&\hspace{2cm}\times\left(c_{\mathbf{k}_{\perp}, 1}^{(\pm)}\psi^{(\pm), \text{reg}}_{\mathbf{k}_{\perp}, 1 }(\mathbf{x}',\omega)+
c_{\mathbf{k}_{\perp}, 2}^{(\pm)} \psi^{(\pm), \text{reg}}_{\mathbf{k}_{\perp}, 2}(\mathbf{x}',\omega)+d_{\mathbf{k}_{\perp},1}^{(\pm)}\psi^{(\pm), \text{out}}_{\mathbf{k}_{\perp}, 1 }(\mathbf{x}',\omega)+d_{\mathbf{k}_{\perp}, 2}^{(\pm) }
\psi^{(\pm), \text{out}}_{\mathbf{k}_{\perp}, 2 }(\mathbf{x}',\omega)\right).\end{split}
\end{equation*}
Here $\mathbf{x}'=\mathbf{x}-\mathbf{L}$, $\mathbf{L}=L\mathbf{e}_z$. The two representations are related by translation matrices $\mathbb{V}$ and $\mathbb{W}$:
\begin{align*}
\psi^{(\pm), \text{reg}}_{\mathbf{k}_{\perp}, 1 }(\mathbf{x}',\omega)=&\sum_{j=\frac{1}{2},\frac{3}{2},\ldots}\sum_{m=-j, -j+1, \ldots, j-1, j}\left(
V^{(\pm), 11}_{jm, \mathbf{k}_{\perp}}\psi_{jm,1}^{(\pm),\text{reg}}(\mathbf{x},\omega)+V^{(\pm), 21}_{jm, \mathbf{k}_{\perp}}\psi_{jm,2}^{(\pm),\text{reg}}(\mathbf{x},\omega)\right),\\
\psi^{(\pm), \text{reg}}_{\mathbf{k}_{\perp}, 2 }(\mathbf{x}',\omega)=&\sum_{j=\frac{1}{2},\frac{3}{2},\ldots}\sum_{m=-j, -j+1, \ldots, j-1, j}\left(
V^{(\pm), 12}_{jm, \mathbf{k}_{\perp}}\psi_{jm,1}^{(\pm),\text{reg}}(\mathbf{x},\omega)+V^{(\pm), 22}_{jm, \mathbf{k}_{\perp}}\psi_{jm,2}^{(\pm),\text{reg}}(\mathbf{x},\omega)\right),\\
\psi^{(\pm), \text{out}}_{jm, 1 }(\mathbf{x},\omega)=&H^2\int_{-\infty}^{\infty}\frac{dk_x}{2\pi}\int_{-\infty}^{\infty}\frac{dk_y}{2\pi} \left(
W^{(\pm), 11}_{ \mathbf{k}_{\perp}, jm}\psi_{\mathbf{k}_{\perp},1}^{(\pm),\text{out}}(\mathbf{x}',\omega)+W^{(\pm), 21}_{\mathbf{k}_{\perp}, jm}\psi_{\mathbf{k}_{\perp},2}^{(\pm),\text{out}}(\mathbf{x}',\omega)\right),\\
\psi^{(\pm), \text{out}}_{jm, 2 }(\mathbf{x},\omega)=&H^2\int_{-\infty}^{\infty}\frac{dk_x}{2\pi}\int_{-\infty}^{\infty}\frac{dk_y}{2\pi} \left(
W^{(\pm), 12}_{ \mathbf{k}_{\perp}, jm}\psi_{\mathbf{k}_{\perp},1}^{(\pm),\text{out}}(\mathbf{x}',\omega)+W^{(\pm), 22}_{\mathbf{k}_{\perp}, jm}\psi_{\mathbf{k}_{\perp},2}^{(\pm),\text{out}}(\mathbf{x}',\omega)\right).
\end{align*}
Using the above expressions, we obtain the following relations
\begin{equation*}\begin{split}
\begin{pmatrix}
a_{jm,1}^{(\pm)}\\a_{jm, 2}^{(\pm)}
\end{pmatrix}=&H^2\int_{-\infty}^{\infty}\frac{dk_x}{2\pi}\int_{-\infty}^{\infty}\frac{dk_y}{2\pi}\begin{pmatrix}V_{jm, \mathbf{k}_{\perp}}^{(\pm), 11}  & V_{ jm, \mathbf{k}_{\perp}}^{(\pm), 12}  \\
 V_{ jm, \mathbf{k}_{\perp}}^{(\pm),21}  & V_{jm, \mathbf{k}_{\perp}}^{(\pm), 22}  \end{pmatrix}\begin{pmatrix}
c_{\mathbf{k}_{\perp},1}^{(\pm)}\\c_{\mathbf{k}_{\perp}, 2}^{(\pm)}
\end{pmatrix},\\
\begin{pmatrix}
d_{\mathbf{k}_{\perp},1}^{(\pm)}\\d_{\mathbf{k}_{\perp},1}^{(\pm)}
\end{pmatrix}=&\sum_{j=\frac{1}{2},\frac{3}{2},\ldots}\sum_{m=-j, -j+1, \ldots, j-1, j}\begin{pmatrix}W_{\mathbf{k}_{\perp}, lm}^{(\pm), 11}  & W_{\mathbf{k}_{\perp}, lm}^{(\pm), 12}  \\
W_{\mathbf{k}_{\perp}, lm}^{(\pm), 21}  & W_{\mathbf{k}_{\perp}, lm}^{(\pm), 22} \end{pmatrix}\begin{pmatrix}
b_{lm, 1}^{(\pm)}\\b_{lm, 2}^{(\pm)}
\end{pmatrix}.\end{split}
\end{equation*}
Matching the boundary conditions on the sphere gives
\begin{align*}
\begin{pmatrix}
b_{jm, 1}^{(\pm)}\\b_{jm, 2}^{(\pm)}
\end{pmatrix}=-\mathbb{T}_{jm}^{(\pm)}\begin{pmatrix}
a_{jm, 1}^{(\pm)}\\a_{jm, 2}^{(\pm)}
\end{pmatrix},
\end{align*}
while solving the boundary conditions on the plane gives
\begin{align*}
\begin{pmatrix}
c_{\mathbf{k}_{\perp}, 1}^{(\pm)}\\c_{\mathbf{k}_{\perp},2}^{(\pm)}
\end{pmatrix}=- {\mathbb{T}}_{\mathbf{k}_{\perp}}^{(\pm)}\begin{pmatrix}
d_{\mathbf{k}_{\perp},1}^{(\pm)}\\d_{\mathbf{k}_{\perp},2}^{(\pm)}
\end{pmatrix}.
\end{align*}
These enter the Casimir energy as written in the main text, formula (1):
\begin{align}
E_{\text{Cas}}
=-\frac{\hbar}{2\pi}\int_0^{\infty} d\xi \sum_{+, -}\text{Tr}\,\ln  \left(\mathbb{I}-\mathbb{N}^{(\pm)}(i\xi)\right),
\nonumber
\end{align}
where
\begin{align*}
N_{jm, j'm'}^{(\pm)}=H^2\int_{-\infty}^{\infty}\frac{dk_x}{2\pi}\int_{-\infty}^{\infty}\frac{dk_y}{2\pi}\mathbb{T}_{jm}^{(\pm)}
\begin{pmatrix}V_{jm, \mathbf{k}_{\perp}}^{(\pm), 11}  & V_{ jm, \mathbf{k}_{\perp}}^{(\pm), 12}  \\
 V_{ jm, \mathbf{k}_{\perp}}^{(\pm), 21}  & V_{jm, \mathbf{k}_{\perp}}^{(\pm), 22}  \end{pmatrix}
 {\mathbb{T}}_{\mathbf{k}_{\perp}}^{(\pm)}
 \begin{pmatrix}W_{\mathbf{k}_{\perp}, j'm'}^{(\pm), 11}  & W_{\mathbf{k}_{\perp}, j'm'}^{(\pm), 12}  \\
W_{\mathbf{k}_{\perp}, j'm'}^{(\pm), 21}  & W_{\mathbf{k}_{\perp}, j'm'}^{(\pm), 22} \end{pmatrix},
\end{align*}
where the following correspondence
\begin{align*}
\mathbb{T}^1 =& \bigl[\mathbb{T}_{jm}\bigr],~~~
\mathbb{T}^2 = \bigl[{\mathbb{T}}_{\mathbf{k}_{\perp}}\bigr],~~~
\mathbb{G}^{12} = \left[ \begin{pmatrix}V_{jm, \mathbf{k}_{\perp}}^{11}  & V_{ jm, \mathbf{k}_{\perp}}^{12}  \\
 V_{ jm, \mathbf{k}_{\perp}}^{21}  & V_{jm, \mathbf{k}_{\perp}}^{22}  \end{pmatrix}\right],
~~~
\mathbb{G}^{21} = \left[\begin{pmatrix}W_{\mathbf{k}_{\perp}, j'm'}^{11}  & W_{\mathbf{k}_{\perp}, j'm'}^{12}  \\
W_{\mathbf{k}_{\perp}, j'm'}^{21}  & W_{\mathbf{k}_{\perp}, j'm'}^{22} \end{pmatrix}\right]
\end{align*}
is understood.

\section{II. Transition Matrices}
The first task is now to use the actual boundary conditions,
\begin{equation*}
(1+i\gamma^{\mu}n_{\mu})\psi\Bigr|_{\text{boundary}}=0,
\end{equation*}
to derive an explicit expressions for the matrices $\mathbb{T}_{jm}$ and ${\mathbb{T}}_{\mathbf{k}_{\perp}}$.

On the exterior of the sphere, using
\begin{align*}
\left.\left(1+i\gamma^{\bar{r}}\right)\psi^{(\pm)}\right|_{r=R}=0 ,
\end{align*}
gives
\begin{align*}
a_{jm, 1}^{(\pm)}\mathcal{C}^{\text{reg}}_j\left(J_j(kR)\mp J_{j+1}(kR)\right)+b_{jm, 1}^{(\pm)}\mathcal{C}^{\text{out}}_j\left(H^{(1)}_j(kR)\mp H^{(1)}_{j+1}(kR)\right)=0,\\
a_{jm,2}^{(\pm)}\mathcal{C}^{\text{reg}}_j\left(J_{j+1}(kR)\pm J_{j}(kR)\right)+b_{jm,2}^{(\pm)}\mathcal{C}^{\text{out}}_j\left(H^{(1)}_{j+1}(kR)\pm H^{(1)}_{j}(kR)\right)=0,
\end{align*}
$$\mathcal{C}_j^{\text{reg}}=i^{-j+\frac{1}{2}},\hspace{1cm}\mathcal{C}_j^{\text{out}}=\frac{\pi  }{2}i^{j+\frac{3}{2}},$$
from which we obtain
\begin{align*}
\mathbb{T}_{jm}^{(\pm)}=\begin{pmatrix} T_{jm}^{(\pm),1}&0\\0&T_{jm}^{(\pm), 2}\end{pmatrix},
\end{align*}
\begin{align*}
T_{jm}^{(\pm),1}=\frac{I_j(\kappa R)\mp iI_{j+1}(\kappa R)}{K_j(\kappa R)\pm iK_{j+1}(\kappa R)},\\
T_{jm}^{(\pm),2}=\frac{I_j(\kappa R)\pm iI_{j+1}(\kappa R)}{K_j(\kappa R)\mp iK_{j+1}(\kappa R)},
\end{align*}
which gives formula (6) in the main text.
At the plane,
\begin{align*}
\left(1-i\gamma^{\bar{3}}\right)\psi^{(\pm )}=0,
\end{align*}
using
\begin{align*}
c_{\mathbf{k}_{\perp},1 }^{(\pm)}\psi^{(\pm ), \text{reg}}_{\mathbf{k}_{\perp}, 1 }(\mathbf{x}',\omega)+
c_{\mathbf{k}_{\perp},2 }^{(\pm)} \psi^{(\pm ), \text{reg}}_{\mathbf{k}_{\perp}, 2}(\mathbf{x}',\omega)=&
\begin{pmatrix}  C^{(\pm)}\\\pm \frac{k_1\sigma_1+k_2\sigma_2-k_3\sigma_3}{k}C^{(\pm)} \end{pmatrix} e^{ik_1x+ik_2y-ik_3z'\mp i\omega t},
\\d_{\mathbf{k}_{\perp},1 }^{(\pm)}\psi^{(\pm), \text{out}}_{\mathbf{k}_{\perp}, 1 }(\mathbf{x}',\omega)+d_{\mathbf{k}_{\perp},2 }^{(\pm)}
\psi^{(\pm), \text{out}}_{\mathbf{k}_{\perp}, 2 }(\mathbf{x}',\omega)=&\begin{pmatrix} D^{(\pm)} \\\pm \frac{k_1\sigma_1+k_2\sigma_2+k_3\sigma_3}{k}D^{(\pm)}\end{pmatrix} e^{ik_1x+ik_2y+ik_3z'\mp i\omega t},
\end{align*}
with
$$C^{(\pm)}=\begin{pmatrix}c_{\mathbf{k}_{\perp},1 }^{(\pm)}\\c_{\mathbf{k}_{\perp},2 }^{(\pm)}\end{pmatrix},\hspace{1cm} D^{(\pm)}=\begin{pmatrix}d_{\mathbf{k}_{\perp},1 }^{(\pm}\\d_{\mathbf{k}_{\perp},2 }^{(\pm)}\end{pmatrix},\hspace{1cm}k_3=\sqrt{k^2-k_{\perp}^2},$$
we get ($\sigma_i$ are the Pauli matrices)
\begin{align*}
\left(1\pm\frac{ik_3}{k}\mp\frac{i\sigma_3\left(k_1\sigma_1+k_2\sigma_2 \right)}{k}\right)C^{(\pm)}=&-\left(1\mp \frac{ik_3}{k}\mp\frac{i\sigma_3\left(k_1\sigma_1+k_2\sigma_2 \right)}{k}\right)D^{(\pm)},
\end{align*}
from which we can obtain
\begin{align*}
\mathbb{T}_{\mathbf{k}_{\perp}}^{(\pm)}
=&\begin{pmatrix} T_{\mathbf{k}_{\perp}}^{(\pm),11}& T_{\mathbf{k}_{\perp}}^{(\pm),12}\\T_{\mathbf{k}_{\perp}}^{(\pm),21}&T_{\mathbf{k}_{\perp}}^{(\pm),22}\end{pmatrix}
= \pm\left(\frac{\sqrt{k^2-k_{\perp}^2}}{ik}+\frac{k_1\sigma_2-k_2\sigma_1}{k}\right),
\end{align*}
or explicitly
\begin{align*}
T_{\mathbf{k}_{\perp}}^{11}=T_{\mathbf{k}_{\perp}}^{22}=&\pm\frac{\sqrt{k^2-k_{\perp}^2}}{ik},\\
T_{\mathbf{k}_{\perp}}^{12}=&\mp \frac{i(k_1-ik_2) }{k},\\
T_{\mathbf{k}_{\perp}}^{21}=& \pm\frac{i(k_1+ik_2) }{k }.
\end{align*}
The above expressions reproduce formula (7) in the main text.

\section{III. Translation Matrices}
\subsection{Matrix ${\mathbb G}^{21}$}
The much more tedious task is to find the translation matrices ${\mathbb G}^{12}$ and ${\mathbb G}^{21}$, which will be explained in the present section.
In the following we will use
$$
\mathbf{k}=  k_x\mathbf{e}_x +k_y\mathbf{e}_y+ k_z\mathbf{e}_z,~~~\mathbf{r}=x\mathbf{e}_x+y\mathbf{e}_y+z\mathbf{e}_z,
$$
with
$k_x=k\sin\theta_k\cos\varphi_k, k_y=k\sin\theta_k\sin\varphi_k, k_z=k\cos\theta_k$.

The first step of our procedure consists in defining the following differential operator $\mathcal{P}_{lm}$
\begin{equation*}
\begin{split}
\mathcal{P}_{lm}=&(-1)^m \sqrt{\frac{2l+1}{4\pi}\frac{(l-m)!}{(l+m)!}}\left(\frac{\pa_x+i\pa_y}{ik}\right)^mP_l^{(m)}\left(\frac{\pa_z}{ik}\right), \\
\mathcal{P}_{l,-m}=&  \sqrt{\frac{2l+1}{4\pi}\frac{(l-m)!}{(l+m)!}}\left(\frac{\pa_x-i\pa_y}{ik}\right)^mP_l^{(m)}\left(\frac{\pa_z}{ik}\right).
\end{split}
\end{equation*}One has
\begin{equation*}
\mathcal{P}_{lm}e^{i\mathbf{k}\cdot\mathbf{r}}=Y_{lm}(\theta_k,\varphi_k)e^{i\mathbf{k}\cdot\mathbf{r}},
\end{equation*}
\begin{equation*}\label{eq5_2_1}
\begin{split}
\mathcal{P}_{lm}j_0(kr)=&i^lj_l(kr)Y_{lm}(\theta,\varphi),\hspace{1cm}\mathcal{P}_{lm}h_0^{(1)}(kr)=i^lh_l^{(1)}(kr)Y_{lm}(\theta,\varphi),\\
j_l(z)=&\sqrt{\frac{\pi}{2z}}J_{l+\frac{1}{2}}(z),
\hspace{2.3cm} h_l^{(1)}(z)=\sqrt{\frac{\pi}{2z}}H_{l+\frac{1}{2}}^{(1)}(z),
\end{split}
\end{equation*}
which allows to express the mode functions given in formula (2) of the main text as
\begin{align*}
\psi_{jm,1}^{(\pm), *}=
\mathcal{C}_j^{*} e^{\mp i\omega t} i^{-j+\frac{1}{2}}\left(\begin{aligned} \sqrt{\frac{j+m}{2j}}\mathcal{P}_{j-\frac{1}{2},m-\frac{1}{2}}\\\sqrt{\frac{j-m}{2j}}\mathcal{P}_{j-\frac{1}{2},m+\frac{1}{2}}\\
\pm\sqrt{\frac{j-m+1}{2j+2}}\mathcal{P}_{j+\frac{1}{2},m-\frac{1}{2}}\\\mp\sqrt{\frac{j+m+1}{2j+2}}\mathcal{P}_{j+\frac{1}{2},m+\frac{1}{2}}\end{aligned}\right)f^*_0(kr),\\
\psi_{jm,2}^{(\pm), *}=
-\mathcal{C}_j^{*} e^{\mp i\omega t} i^{-j-\frac{1}{2}}\left(\begin{aligned}
\sqrt{\frac{j-m+1}{2j+2}}\mathcal{P}_{j+\frac{1}{2},m-\frac{1}{2}}\\-\sqrt{\frac{j+m+1}{2j+2}}\mathcal{P}_{j+\frac{1}{2},m+\frac{1}{2}}\\
\pm\sqrt{\frac{j+m}{2j}}\mathcal{P}_{j-\frac{1}{2},m-\frac{1}{2}}\\\pm\sqrt{\frac{j-m}{2j}}\mathcal{P}_{j-\frac{1}{2},m+\frac{1}{2}}\end{aligned}\right)f_0^*(kr),
\end{align*}
where $f_l^{*}(k r)$ is defined in formula (3) of the main text.
We may now use the following integral representation
\begin{equation*}
\begin{split}
h_0(kr)=&\frac{\exp(ikr)}{ikr}=\frac{1}{2\pi}\int_{-\infty}^{\infty} dk_x\int_{-\infty}^{\infty} dk_y \frac{e^{ik_xx+ik_yy \pm i\sqrt{k^2-k_x^2-k_y^2}z}}
{k\sqrt{k^2-k_x^2-k_y^2}},\quad z\gtrless 0,
\end{split}
\end{equation*}
and express the spinors $\psi_{jm,1}^{(\pm ), \text{out}}(\mathbf{x},\omega)$, after some calculation, as
\begin{align*}
\psi_{jm,1}^{(\pm ), \text{out}}(\mathbf{x},\omega)=
&\mathcal{C}_j^{\text{out}} e^{\mp i\omega t} i^{-j+\frac{1}{2}}\frac{1}{4\pi^{\frac{3}{2}}}\sqrt{\frac{(j-m)!}{(j+m)!}}\int_{-\infty}^{\infty} dk_x\int_{-\infty}^{\infty} dk_ye^{i\left(m-\frac{1}{2}\right)\varphi_k}
\left(\begin{aligned} (j+m)P_{j-\frac{1}{2}}^{m-\frac{1}{2}}(\cos\theta_k)\\ P_{j-\frac{1}{2}}^{m+\frac{1}{2}}
(\cos\theta_k)e^{i \varphi_k}\\
\pm (j-m+1)P_{j+\frac{1}{2}}^{m-\frac{1}{2}}(\cos\theta_k) \\
\mp P_{j+\frac{1}{2}}^{m+\frac{1}{2}}(\cos\theta_k)e^{i \varphi_k}\end{aligned}\right)\\&\times \frac{e^{ik_xx+ik_yy + i\sqrt{k^2-k_x^2-k_y^2}z}}
{k\sqrt{k^2-k_x^2-k_y^2}}.
\end{align*}
Using the definitions of $\theta_k$ and $\varphi_k$ given at the beginning of the section, one finds
\begin{align*}
&\left(\begin{aligned} (j+m)P_{j-\frac{1}{2}}^{m-\frac{1}{2}}(\cos\theta_k)\\ P_{j-\frac{1}{2}}^{m+\frac{1}{2}}
(\cos\theta_k)e^{i \varphi_k}\\
\pm(j-m+1)P_{j+\frac{1}{2}}^{m-\frac{1}{2}}(\cos\theta_k) \\
\mp P_{j+\frac{1}{2}}^{m+\frac{1}{2}}(\cos\theta_k)e^{i \varphi_k}\end{aligned}\right)
 =
 (j+m)P_{j-\frac{1}{2}}^{m-\frac{1}{2}}(\cos\theta_k)\left(\begin{aligned} 1\hspace{0.5cm}\\0 \hspace{0.5cm}\\ \pm\frac{\sqrt{k^2-k_{\perp}^2}}{k}\\\pm\frac{k_1+ik_2}{k}\end{aligned}\right)+P_{j-\frac{1}{2}}^{m+\frac{1}{2}}
(\cos\theta_k)e^{i \varphi_k}\left(\begin{aligned} 0\hspace{0.5cm}\\1 \hspace{0.5cm}\\ \pm\frac{k_1-ik_2}{k}\\\mp\frac{\sqrt{k^2-k_{\perp}^2}}{k}\end{aligned}\right).
\end{align*}
Recalling that
\begin{align*}
\psi^{(\pm ), \text{out}}_{jm, 1 }(\mathbf{x},\omega)=&H^2\int_{-\infty}^{\infty}\frac{dk_x}{2\pi}\int_{-\infty}^{\infty}\frac{dk_y}{2\pi} \left(
W^{(\pm ), 11}_{ \mathbf{k}_{\perp}, jm}\psi_{\mathbf{k}_{\perp},1}^{(\pm ),\text{out}}(\mathbf{x}',\omega)+W^{(\pm ), 21}_{\mathbf{k}_{\perp}, jm}\psi_{\mathbf{k}_{\perp},2}^{(\pm ),\text{out}}(\mathbf{x}',\omega)\right),
\end{align*}
and using the explicit expression for the plane waves (formulae (4) - (5) in the main text), we arrive at
\begin{align*}
W^{(\pm ), 11}_{ \mathbf{k}_{\perp}, jm}=&-\frac{\pi^{\frac{3}{2}}}{2H^2}\sqrt{\frac{(j-m)!}{(j+m)!}}\frac{1}{k\sqrt{k^2-k_{\perp}^2}}e^{i\left(m-\frac{1}{2}\right)\varphi_k}(j+m)P_{j-\frac{1}{2}}^{m-\frac{1}{2}}(\cos\theta_k)
e^{i\sqrt{k^2-k_{\perp}^2}L},\\
W^{(\pm ), 21}_{ \mathbf{k}_{\perp}, jm}=&-\frac{\pi^{\frac{3}{2}}}{2H^2}\sqrt{\frac{(j-m)!}{(j+m)!}}\frac{1}{k\sqrt{k^2-k_{\perp}^2}}e^{i\left(m+\frac{1}{2}\right)\varphi_k} P_{j-\frac{1}{2}}^{m+\frac{1}{2}}(\cos\theta_k)e^{i\sqrt{k^2-k_{\perp}^2}L}.
\end{align*}
Obtaining $W^{(\pm ), 12}_{ \mathbf{k}_{\perp}, jm}$ and $W^{(\pm ), 22}_{ \mathbf{k}_{\perp}, jm}$ follows from similar steps, leading to
\begin{align*}
W^{(\pm ), 12}_{ \mathbf{k}_{\perp}, jm}=&-\frac{i\pi^{\frac{3}{2}}}{2H^2}\sqrt{\frac{(j-m)!}{(j+m)!}}\frac{1}{k\sqrt{k^2-k_{\perp}^2}}e^{i\left(m-\frac{1}{2}\right)\varphi_k}(j-m+1)P_{j+\frac{1}{2}}^{m-\frac{1}{2}}
(\cos\theta_k),\\
W^{(\pm ), 22}_{ \mathbf{k}_{\perp}, jm}=&\frac{i\pi^{\frac{3}{2}}}{2H^2}\sqrt{\frac{(j-m)!}{(j+m)!}}\frac{1}{k\sqrt{k^2-k_{\perp}^2}}e^{i\left(m+\frac{1}{2}\right)\varphi_k} P_{j+\frac{1}{2}}^{m+\frac{1}{2}}(\cos\theta_k).
\end{align*}
Combining all the $W^{(\pm ), ij}_{ \mathbf{k}_{\perp}, jm}$ gives
\begin{align*}
{\mathbb G}^{21,(\pm)}_{ \mathbf{k}_{\perp}, jm} =& \begin{pmatrix} W^{(\pm),11}_{ \mathbf{k}_{\perp}, jm}&W^{(\pm),12}_{ \mathbf{k}_{\perp}, jm}\\W^{(\pm),21}_{ \mathbf{k}_{\perp}, jm}&W^{(\pm),22}_{ \mathbf{k}_{\perp}, jm}\end{pmatrix}=\\
=&
-\frac{\pi^{\frac{3}{2}}}{2H^2}\sqrt{\frac{(j-m)!}{(j+m)!}}\frac{1}{k\sqrt{k^2-k_{\perp}^2}}e^{i\left(m-\frac{1}{2}\right)\varphi_k}
\begin{pmatrix}(j+m)P_{j-\frac{1}{2}}^{m-\frac{1}{2}}(\cos\theta_k)&i(j-m+1)P_{j+\frac{1}{2},m-\frac{1}{2}}(\cos\theta_k)\\P_{j-\frac{1}{2}}^{m+\frac{1}{2}}(\cos\theta_k)e^{i\varphi_k}&
-iP_{j+\frac{1}{2},m+\frac{1}{2}}(\cos\theta_k)e^{i\varphi_k}
\end{pmatrix},
\end{align*}
as presented in the main text in formula (10).

\subsection{Matrix ${\mathbb G}^{12}$}
In order to find ${\mathbb G}^{12}$, we may proceed as follows. First of all, we use the integral representation
\begin{equation*}
\begin{split}
j_0(kr)=&\frac{\sin (kr)}{kr}=\frac{1}{4\pi}\int_0^{2\pi}d\varphi_k\int_0^{\pi}d\theta_k \sin\theta_k e^{i\mathbf{k}\cdot\mathbf{r}}\end{split}\end{equation*}
to express the regular spherical solutions as
\begin{align*}
\psi_{jm,1}^{(\pm), \text{reg}}(\mathbf{x},\omega)=\mathcal{C}_j^{\text{reg}} e^{\mp i\omega t} i^{-j+\frac{1}{2}}\frac{1}{8\pi^{\frac{3}{2}}}\sqrt{\frac{(j-m)!}{(j+m)!}}\int_0^{2\pi}d\varphi_k\int_0^{\pi}d\theta_k \sin\theta_k \left(\begin{aligned} (j+m)P_{j-\frac{1}{2}}^{m-\frac{1}{2}}(\cos\theta_k)e^{i\left(m-\frac{1}{2}\right)\varphi_k}\\P_{j-\frac{1}{2}}^{m+\frac{1}{2}}
(\cos\theta_k)e^{i\left(m+\frac{1}{2}\right)\varphi_k}\\
 \pm(j-m+1) P_{j+\frac{1}{2}}^{m-\frac{1}{2}}(\cos\theta_k)e^{i\left(m-\frac{1}{2}\right)\varphi_k}\\
\mp P_{j+\frac{1}{2}}^{m+\frac{1}{2}}(\cos\theta_k)e^{i\left(m+\frac{1}{2}\right)\varphi_k}\end{aligned}\right)e^{i\mathbf{k}\cdot\mathbf{r}},
\end{align*}
\begin{align*}
\psi_{jm,2}^{(\pm), \text{reg}}(\mathbf{x},\omega)=
-\mathcal{C}_j^{\text{reg}} e^{\mp i\omega t} i^{-j-\frac{1}{2}}\frac{1}{8\pi^{\frac{3}{2}}}\sqrt{\frac{(j-m)!}{(j+m)!}}\int_0^{2\pi}d\varphi_k\int_0^{\pi}d\theta_k \sin\theta_k \left(\begin{aligned}
(j-m+1)P_{j+\frac{1}{2}}^{m-\frac{1}{2}}(\cos\theta_k)e^{i\left(m-\frac{1}{2}\right)\varphi_k}\\
-P_{j+\frac{1}{2}}^{m+\frac{1}{2}}(\cos\theta_k)e^{i\left(m+\frac{1}{2}\right)\varphi_k}\\\pm(j+m)P_{j-\frac{1}{2}}^{m-\frac{1}{2}}(\cos\theta_k)e^{i\left(m-\frac{1}{2}\right)\varphi_k}\\
 \pm P_{j-\frac{1}{2}}^{m+\frac{1}{2}}
(\cos\theta_k)e^{i\left(m+\frac{1}{2}\right)\varphi_k}\end{aligned}\right)e^{i\mathbf{k}\cdot\mathbf{r}}.
\end{align*}
We now introduce the following operators
\begin{align*}
\boldsymbol{\mathcal{P}}_{jm,1}^{(\pm)}=\frac{(-1)^{m-\frac{1}{2}}}{2}\left(\begin{aligned} \sqrt{\frac{j+m}{2j}}\mathcal{P}_{j-\frac{1}{2},-m+\frac{1}{2}}\\-\sqrt{\frac{j-m}{2j}}\mathcal{P}_{j-\frac{1}{2},-m-\frac{1}{2}}\\
\pm\sqrt{\frac{j-m+1}{2j+2}}\mathcal{P}_{j+\frac{1}{2},-m+\frac{1}{2}}\\\pm\sqrt{\frac{j+m+1}{2j+2}}\mathcal{P}_{j+\frac{1}{2},-m-\frac{1}{2}}\end{aligned}\right),~~
\boldsymbol{\mathcal{P}}_{jm,2}^{(\pm)}=\frac{(-1)^{m-\frac{1}{2}}}{2}\left(\begin{aligned} \sqrt{\frac{j-m+1}{2j+2}}\mathcal{P}_{j+\frac{1}{2},-m+\frac{1}{2}}\\\sqrt{\frac{j+m+1}{2j+2}}\mathcal{P}_{j+\frac{1}{2},-m-\frac{1}{2}}\\ \pm\sqrt{\frac{j+m}{2j}}\mathcal{P}_{j-\frac{1}{2},-m+\frac{1}{2}}\\\mp\sqrt{\frac{j-m}{2j}}\mathcal{P}_{j-\frac{1}{2},-m-\frac{1}{2}}\end{aligned}\right),
\end{align*}
that satisfy
\begin{align*}
\boldsymbol{\mathcal{P}}_{j'm',1}^{(\pm)}\cdot \psi_{jm,1}^{(\pm), \text{reg}}(\mathbf{x},\omega)\Biggr|_{\mathbf{x}=0}=&(-1)^{-j+\frac{1}{2}}\frac{e^{\mp i\omega t}}{4\pi}\delta_{jj'}\delta_{mm'},\\
\boldsymbol{\mathcal{P}}_{j'm',2}^{(\pm)}\cdot \psi_{jm,1}^{(\pm), \text{reg}}(\mathbf{x},\omega)\Biggr|_{\mathbf{x}=0}=&0,\\
\boldsymbol{\mathcal{P}}_{j'm',1}^{(\pm)}\cdot \psi_{jm,2}^{(\pm), \text{reg}}(\mathbf{x},\omega)\Biggr|_{\mathbf{x}=0}=&0\\
\boldsymbol{\mathcal{P}}_{j'm',2}^{(\pm)}\cdot \psi_{jm,2}^{(\pm), \text{reg}}(\mathbf{x},\omega)\Biggr|_{\mathbf{x}=0}=&(-1)^{-j+\frac{1}{2}}i\frac{e^{\mp i\omega t}}{4\pi}\delta_{jj'}\delta_{mm'}.
\end{align*}
Recalling that
\begin{align*}
\psi^{(\pm), \text{reg}}_{\mathbf{k}_{\perp}, 1 }(\mathbf{x}',\omega)=&\sum_{j=\frac{1}{2},\frac{3}{2},\ldots}\sum_{m=-j, -j+1, \ldots, j-1, j}\left(
V^{(\pm),11}_{jm, \mathbf{k}_{\perp}}\psi_{jm,1}^{(\pm),\text{reg}}(\mathbf{x},\omega)+V^{(\pm),21}_{jm, \mathbf{k}_{\perp}}\psi_{jm,2}^{(\pm),\text{reg}}(\mathbf{x},\omega)\right),\\
\psi^{(\pm), \text{reg}}_{\mathbf{k}_{\perp}, 2 }(\mathbf{x}',\omega)=&\sum_{j=\frac{1}{2},\frac{3}{2},\ldots}\sum_{m=-j, -j+1, \ldots, j-1, j}\left(
V^{(\pm),12}_{jm, \mathbf{k}_{\perp}}\psi_{jm,1}^{(\pm),\text{reg}}(\mathbf{x},\omega)+V^{(\pm),22}_{jm, \mathbf{k}_{\perp}}\psi_{jm,2}^{(\pm),\text{reg}}(\mathbf{x},\omega)\right),
\end{align*}
we can use the above relations for the operators $\boldsymbol{\mathcal{P}}_{jm,i}^{(\pm)}$ to extract the matrix elements $V^{(\pm),ij}_{jm, \mathbf{k}_{\perp}}$:
\begin{align*}
V^{(\pm),11}_{jm, \mathbf{k}_{\perp}}=&(-1)^{-j+\frac{1}{2}}4\pi e^{\pm i\omega t}\boldsymbol{\mathcal{P}}^{(\pm)}_{jm,1} \cdot \psi^{(\pm), \text{reg}}_{\mathbf{k}_{\perp}, 1 }(\mathbf{x}-\mathbf{L},\omega)\Biggr|_{\mathbf{x}=0}\\
=&(-1)^{-m+\frac{1}{2}}\sqrt{4\pi}  \sqrt{\frac{(j-m)!}{(j+m)!}}\left(j+m\right)P_{j-\frac{1}{2}}^{m-\frac{1}{2}}\left(\cos\theta_k\right)e^{-i\left(m-\frac{1}{2}\right)\varphi_k}
e^{i\sqrt{k^2-k_{\perp}^2}L},\\
V^{(\pm)21}_{jm, \mathbf{k}_{\perp}}=&-(-1)^{-j+\frac{1}{2}}4\pi i e^{\pm i\omega t}\boldsymbol{\mathcal{P}}^{(\pm)}_{jm,2} \cdot \psi^{(\pm), \text{reg}}_{\mathbf{k}_{\perp}, 1 }(\mathbf{x}-\mathbf{L},\omega)\Biggr|_{\mathbf{x}=0}\\
=&(-1)^{-m+\frac{1}{2}}\sqrt{4\pi}  i\sqrt{\frac{(j-m)!}{(j+m)!}}\left(j-m+1\right)P_{j+\frac{1}{2}}^{m-\frac{1}{2}}\left(\cos\theta_k\right)e^{-i\left(m-\frac{1}{2}\right)\varphi_k}e^{i\sqrt{k^2-k_{\perp}^2}L},\\
V^{(\pm),12}_{jm, \mathbf{k}_{\perp}}=&(-1)^{-j+\frac{1}{2}}4\pi e^{\pm i\omega t}\boldsymbol{\mathcal{P}}_{jm,1}^{(\pm)}\cdot \psi^{(\pm), \text{reg}}_{\mathbf{k}_{\perp}, 2 }(\mathbf{x}-\mathbf{L},\omega)\Biggr|_{\mathbf{x}=0}\\
=&(-1)^{-m-\frac{1}{2}}\sqrt{4\pi}  \sqrt{\frac{(j-m)!}{(j+m)!}} P_{j-\frac{1}{2}}^{m+\frac{1}{2}}\left(\cos\theta_k\right)e^{-i\left(m+\frac{1}{2}\right)\varphi_k}e^{i\sqrt{k^2-k_{\perp}^2}L},\\
V^{(\pm),22}_{jm, \mathbf{k}_{\perp}}=&-(-1)^{-j+\frac{1}{2}}4\pi ie^{\pm i\omega t}\boldsymbol{\mathcal{P}}_{jm,2}^{(\pm)}\cdot \psi^{(+), \text{reg}}_{\mathbf{k}_{\perp}, 2 }(\mathbf{x}-\mathbf{L},\omega)\Biggr|_{\mathbf{x}=0}\\
=&(-1)^{-m+\frac{1}{2}}\sqrt{4\pi} i \sqrt{\frac{(j-m)!}{(j+m)!}} P_{j+\frac{1}{2}}^{m+\frac{1}{2}}\left(\cos\theta_k\right)e^{-i\left(m+\frac{1}{2}\right)\varphi_k}e^{i\sqrt{k^2-k_{\perp}^2}L}.
\end{align*}
Combining everything we arrive at
\begin{align*}
{\mathbb G}^{(\pm),12}_{jm,{\bf k}_\perp} =&
\begin{pmatrix}
V^{(\pm),11}_{ \mathbf{k}_{\perp}, jm}&V^{(\pm),12}_{ \mathbf{k}_{\perp}, jm}\\V^{(\pm),21}_{ \mathbf{k}_{\perp}, jm}&V^{(\pm),22}_{ \mathbf{k}_{\perp}, jm}
\end{pmatrix}
=\\
=&(-1)^{-m+\frac{1}{2}}
\sqrt{4\pi}\sqrt{\frac{(j-m)!}{(j+m)!}} e^{-i\left(m-\frac{1}{2}\right)\varphi_k}
\begin{pmatrix}(j+m)P_{j-\frac{1}{2}}^{m-\frac{1}{2}}(\cos\theta_k)&-P_{j-\frac{1}{2}}^{m+\frac{1}{2}}(\cos\theta_k)e^{-i\varphi_k}\\i(j-m+1)P_{j+\frac{1}{2}}^{m-\frac{1}{2}}(\cos\theta_k)&
iP_{j+\frac{1}{2}}^{m+\frac{1}{2}}(\cos\theta_k)e^{-i\varphi_k}
\end{pmatrix},
\end{align*}
as reported in formula (9) of the main text.

\section{IV. Matrix $\mathbb N$}

The matrix $\mathbb N$ can be obtained by combining the translation and transition matrices leading to
\begin{align*}
N_{jm, j'm'}^{\pm}=&\pm (-1)^{-m+\frac{1}{2}}\sqrt{\frac{(j-m)!(j'-m)!}{(j+m)!(j'+m)!}}\left(\begin{aligned}\frac{I_j(\kappa R)\mp iI_{j+1}(\kappa R)}{K_j(\kappa R)\pm iK_{j+1}(\kappa R)} & \hspace{2cm}0\hspace{1cm}\\\hspace{1cm}0\hspace{2cm}
 &\frac{I_j(\kappa R)\pm iI_{j+1}(\kappa R)}{K_j(\kappa R)\mp iK_{j+1}(\kappa R)} \end{aligned}\right) \times
 \\&
 \times
 \int_{0}^{\infty}dk_{\perp}\,k_{\perp} \frac{i\pi\delta_{m,m'}}{2k\sqrt{k^2-k_{\perp}^2}} \begin{pmatrix}A&B\\C&
D
\end{pmatrix}
e^{2i\sqrt{k^2-k_{\perp}^2}L},
\end{align*}
where
\begin{align*}
A=&(j+m)(j'-m+1)P_{j-\frac{1}{2}}^{m-\frac{1}{2}}(\cos\theta_k)P_{j'+\frac{1}{2}}^{m-\frac{1}{2}}(\cos\theta_k)-P_{j-\frac{1}{2}}^{m+\frac{1}{2}}(\cos\theta_k)
 P_{j'+\frac{1}{2}}^{m+\frac{1}{2}}(\cos\theta_k),\\
 B=&i\left((j+m)(j'+m)P_{j-\frac{1}{2}}^{m-\frac{1}{2}}(\cos\theta_k)
 P_{j'-\frac{1}{2}}^{m-\frac{1}{2}}(\cos\theta_k)+P_{j-\frac{1}{2}}^{m+\frac{1}{2}}(\cos\theta_k) P_{j'-\frac{1}{2}}^{m+\frac{1}{2}}(\cos\theta_k)\right),\\
 C=&i\left((j-m+1)(j'-m+1)P_{j+\frac{1}{2}}^{m-\frac{1}{2}}(\cos\theta_k)P_{j'+\frac{1}{2}}^{m-\frac{1}{2}}(\cos\theta_k)+P_{j+\frac{1}{2}}^{m+\frac{1}{2}}(\cos\theta_k) P_{j'+\frac{1}{2}}^{m+\frac{1}{2}}(\cos\theta_k)\right),\\
 D=&-(j-m+1)(j'+m)P_{j+\frac{1}{2}}^{m-\frac{1}{2}}(\cos\theta_k) P_{j'-\frac{1}{2}}^{m-\frac{1}{2}}(\cos\theta_k)+P_{j+\frac{1}{2}}^{m+\frac{1}{2}}(\cos\theta_k)P_{j'-\frac{1}{2}}^{m+\frac{1}{2}}(\cos\theta_k).
\end{align*}
Using the following relation
\begin{align*}
&(j-m+1)(j'-m+1)P_{j+\frac{1}{2}}^{m-\frac{1}{2}}(\cos\theta_k)P_{j'+\frac{1}{2}}^{m-\frac{1}{2}}(\cos\theta_k)+P_{j+\frac{1}{2}}^{m+\frac{1}{2}}(\cos\theta_k) P_{j'+\frac{1}{2}}^{m+\frac{1}{2}}(\cos\theta_k)\\
=&(j+m)(j'+m)P_{j-\frac{1}{2}}^{m-\frac{1}{2}}(\cos\theta_k)
 P_{j'-\frac{1}{2}}^{m-\frac{1}{2}}(\cos\theta_k)+P_{j-\frac{1}{2}}^{m+\frac{1}{2}}(\cos\theta_k) P_{j'-\frac{1}{2}}^{m+\frac{1}{2}}(\cos\theta_k),
\end{align*}
we can prove that $B=C$. Then, noticing that
\begin{align*}
\begin{pmatrix}A & B\\C &D\end{pmatrix}
=&\begin{pmatrix}
(j+m)P_{j-\frac{1}{2}}^{m-\frac{1}{2}}(\cos\theta_k)&-P_{j-\frac{1}{2}}^{m+\frac{1}{2}}(\cos\theta_k)e^{-i\varphi_k}\\ i(j-m+1)P_{j+\frac{1}{2}}^{m-\frac{1}{2}}(\cos\theta_k)&
iP_{j+\frac{1}{2}}^{m+\frac{1}{2}}(\cos\theta_k)e^{-i\varphi_k}
\end{pmatrix}
\begin{pmatrix}
(j'-m+1)P_{j'+\frac{1}{2}}^{m-\frac{1}{2}}(\cos\theta_k)&i(j'+m)P_{j'-\frac{1}{2}}^{m-\frac{1}{2}}(\cos\theta_k)\\ P_{j'+\frac{1}{2}}^{m+\frac{1}{2}}(\cos\theta_k)e^{i\varphi_k}&
- iP_{j'-\frac{1}{2}}^{m+\frac{1}{2}}(\cos\theta_k)e^{i\varphi_k}
\end{pmatrix},
\end{align*}
we can express $N_{jm, j'm'}^{\pm}$ in terms of
\begin{align*}
\bar{P}_{l}^m(z)=&\frac{(-1)^m}{2^ll!}(z^2-1)^{\frac{m}{2}}\frac{d^{l+m}}{dz^{l+m}}(z^2-1)^l,\quad m\geq 0,\\
\bar{P}_{l}^{-m}(z)=&(-1)^m \frac{(l-m)!}{(l+m)!}\bar{P}_l^m(z),
\end{align*}
as
\begin{align*}
{M}_{jm, j'm'}^{\pm}=&\mp \frac{i\pi\delta_{m,m'}}{2}\sqrt{\frac{(j-m)!(j'-m)!}{(j+m)!(j'+m)!}}\left(\begin{aligned}\frac{I_j(\kappa R)\mp iI_{j+1}(\kappa R)}{K_j(\kappa R)\pm iK_{j+1}(\kappa R)} & \hspace{2cm}0\hspace{1cm}\\\hspace{1cm}0\hspace{2cm}
 &-\frac{I_j(\kappa R)\pm iI_{j+1}(\kappa R)}{K_j(\kappa R)\mp iK_{j+1}(\kappa R)} \end{aligned}\right)
 \int_{0}^{\infty}d\theta\sinh\theta e^{-2\kappa L\cosh\theta}\\&\times \begin{pmatrix}(j+m)\bar{P}_{j-\frac{1}{2}}^{m-\frac{1}{2}}(\cosh\theta)&\bar{P}_{j-\frac{1}{2}}^{m+\frac{1}{2}}(\cosh\theta)\\ (j-m+1)\bar{P}_{j+\frac{1}{2}}^{m-\frac{1}{2}}(\cosh\theta)&
-\bar{P}_{j+\frac{1}{2}}^{m+\frac{1}{2}}(\cosh\theta)
\end{pmatrix}\begin{pmatrix}(j'-m+1)\bar{P}_{j'+\frac{1}{2}}^{m-\frac{1}{2}}(\cosh\theta)&(j'+m)\bar{P}_{j'-\frac{1}{2}}^{m-\frac{1}{2}}(\cosh\theta)\\ \bar{P}_{j'+\frac{1}{2}}^{m+\frac{1}{2}}(\cosh\theta) &-
\bar{P}_{j'-\frac{1}{2}}^{m+\frac{1}{2}}(\cosh\theta)
\end{pmatrix},
\end{align*}
where we have defined
\begin{align*}
{\mathbb M} =
\begin{pmatrix}
1 & 0\\ 0& i
\end{pmatrix}
\mathbb N
\begin{pmatrix}
1 & 0\\ 0& -i
\end{pmatrix}.
\end{align*}
From the above expression, it is straightforward to prove that
\begin{align*}
{M}_{jm, j'm'}^+=\overline{{M}_{jm, j'm'}^-},
\end{align*}
and that
\begin{align*}
{M}_{j,-m; j',-m}^{+} = {M}_{j,m; j',m}^{+},
\end{align*}
from which it follows that
\begin{align}
E_{\text{Cas}}
=-\frac{\hbar c}{\pi}\text{Re}\,\int_0^{\infty} d\kappa \text{Tr}\,\ln  \left(\mathbb{I}-{\mathbb{M}}^+ \right),
\end{align}
from which formulae (11) and (12) in the main text can be obtained straightforwardly.

\section{V. Small separation asymptotic behaviour}

In order to compute the behaviour of the Casimir energy at short distance, we first expand the logarithm
\begin{align*}
E_{\text{Cas}}
=&\frac{\hbar c}{\pi}\sum_{s=0}^{\infty}\frac{1}{s+1}\text{Re}\,\int_0^{\infty} d\kappa \sum_{m=\ldots,-\frac{3}{2},-\frac{1}{2},\frac{1}{2},\frac{3}{2}, \ldots}\sum_{j_0=|m|}^{\infty}\sum_{j_1=|m|}^{\infty} \ldots\sum_{j_s=|m|}^{\infty}
{M}^+_{j_0m,j_1m}  \ldots {M}^+_{j_sm,j_0m},
\end{align*}
and let
\begin{gather*}
j_0=l,\quad j_i\mapsto j+n_i,\\
\vep =\frac{d}{R},\quad \kappa R=\omega=\frac{l\sqrt{1-\tau^2}}{\tau},
\end{gather*}
which allows us to write
\begin{align*}
E_{\text{Cas}}
= &\frac{\hbar c}{\pi R}\sum_{s=0}^{\infty}\frac{1}{s+1}\text{Re}\, \int_0^1\frac{d\tau}{\tau^2\sqrt{1-\tau^2}} \int_0^{\infty} dl\, l \int_{-\infty}^{\infty} dm\int_{-\infty}^{\infty}dn_1\ldots \int_{-\infty}^{\infty} dn_s\;
{M}^+_{l,l+n_1}  \ldots {M}^+_{l+n_s,l}.
\end{align*}
Using the following representations
\begin{align*}
\bar{P}_{j-\frac{1}{2}}^{m-\frac{1}{2}}(\cosh\theta)=&
\frac{(j+m-1)!}{\pi  }\sum_{k=0}^{j-\frac{1}{2}}\frac{1}{k!\left(j-\frac{1}{2}-k\right)!}e^{\left(j-\frac{1}{2}-2k\right)\theta}\int_{-\frac{\pi}{2}}^{\frac{\pi}{2}}d\varphi  \cos^{2j-1-2k}\varphi \sin^{2k}\varphi e^{2i\left(m-\frac{1}{2}\right)\varphi},\\
\bar{P}_{j-\frac{1}{2}}^{m+\frac{1}{2}}(\cosh\theta)
=& \frac{(j+m)!}{\pi  }\sum_{k=0}^{j-\frac{1}{2}}\frac{1}{k!\left(j-\frac{1}{2}-k\right)!}e^{\left(j-\frac{1}{2}-2k\right)\theta}\int_{-\frac{\pi}{2}}^{\frac{\pi}{2}}d\varphi  \cos^{2j-1-2k}\varphi \sin^{2k}\varphi e^{2i\left(m+\frac{1}{2}\right)\varphi},\\
\bar{P}_{j+\frac{1}{2}}^{m-\frac{1}{2}}(\cosh\theta)
=& \frac{(j+m)!}{\pi  }\sum_{k=0}^{j+\frac{1}{2}}\frac{1}{k!\left(j+\frac{1}{2}-k\right)!}e^{\left(j+\frac{1}{2}-2k\right)\theta}\int_{-\frac{\pi}{2}}^{\frac{\pi}{2}}d\varphi  \cos^{2j+1-2k}\varphi \sin^{2k}\varphi e^{2i\left(m-\frac{1}{2}\right)\varphi},\\
\bar{P}_{j+\frac{1}{2}}^{m+\frac{1}{2}}(\cosh\theta)
=& \frac{(j+m+1)!}{\pi  }\sum_{k=0}^{j+\frac{1}{2}}\frac{1}{k!\left(j+\frac{1}{2}-k\right)!}e^{\left(j+\frac{1}{2}-2k\right)\theta}\int_{-\frac{\pi}{2}}^{\frac{\pi}{2}}d\varphi  \cos^{2j+1-2k}\varphi \sin^{2k}\varphi e^{2i\left(m+\frac{1}{2}\right)\varphi},
\end{align*}
we obtain, after some algebra,
\begin{align*}
{M}_{l+n_i, l+n_{i+1}}^{+} = &- \frac{i}{2\pi}\sqrt{ (l+n_i-m)!(l+n_{i+1}-m)!(l+n_i+m)!(l+n_{i+1}+m)!}\\&\times \left(\begin{aligned}\frac{I_{l+n_i}(\omega)- iI_{l+n_i+1}(\omega)}{K_{l+n_i}(\omega)+ iK_{l+n_i+1}(\omega)} & \hspace{2cm}0\hspace{1cm}\\\hspace{1cm}0\hspace{2cm}
 &-\frac{I_{l+n_i}(\omega)+ iI_{l+n_i+1}(\omega)}{K_{l+n_i+1}(\omega)- iK_{l+n_i+1}(\omega)} \end{aligned}\right)
 \int_{0}^{\infty}d\theta\sinh\theta e^{-2(1+\vep)\omega\cosh\theta}\\&\times
 \sum_{k=0}^{\infty}\frac{1}{k!(l+n_i+\frac{1}{2}-k)!}\sum_{k'=0}^{\infty}\frac{1}{k'!(l+n_{i+1}+\frac{1}{2}-k')!}e^{(2l+n_i+n_{i+1}-2k-2k')\theta}
 \\&\times\int_{-\frac{\pi}{2}}^{\frac{\pi}{2}}d\varphi\cos^{2l+2n_i-2k}\varphi\sin^{2k}\varphi e^{2im\varphi}
 \int_{-\frac{\pi}{2}}^{\frac{\pi}{2}}d\varphi'\cos^{2l+2n_{i+1}-2k'}\varphi'\sin^{2k'}\varphi' e^{2im\varphi'}
 \begin{pmatrix}  A & B\\ C & D\end{pmatrix},
\end{align*}
where
\begin{align*}
A=&\left(l+n_i+\frac{1}{2}-k\right)\frac{\cos\varphi'}{\cos\varphi}\left[\left(l+n_{i+1}-m+1\right)e^{-i(\varphi+\varphi')}+\left(l+n_{i+1}+m+1\right)e^{i(\varphi+\varphi')}\right],\\
B=&\left(l+n_i+\frac{1}{2}-k\right)\left(l+n_{i+1}+\frac{1}{2}-k'\right)\frac{e^{-\theta}}{\cos\varphi\cos\varphi'}\left[e^{-i(\varphi+\varphi')}-e^{i(\varphi+\varphi')}\right],\\
C=&e^{\theta}\cos\varphi\cos\varphi'\left[\left(l+n_i-m+1\right)\left(l+n_{i+1}-m+1\right)e^{-i(\varphi+\varphi')}-\left(l+n_i+m+1\right)\left(l+n_{i+1}+m+1\right)e^{i(\varphi+\varphi')}\right],\\
D=&\left(l+n_{i+1}+\frac{1}{2}-k'\right)\frac{\cos\varphi}{\cos\varphi'}\left[\left(l+n_i-m+1\right)e^{-i(\varphi+\varphi')}+\left(l+n_{i}+m+1\right)e^{i(\varphi+\varphi')}\right].
\end{align*}
We then make the following substitutions
\begin{align*}
\theta\mapsto \theta+\theta_0,\quad \text{where}\quad \sinh\theta_0=\frac{\tau}{\sqrt{1-\tau^2}}
\end{align*}
and,   with the aid of a symbolic manipulation program, we expand up to first order in $\varepsilon$ keeping in mind that $l\sim\vep^{-1}$, $n_i, m\sim \vep^{-\frac{1}{2}}$, $ \varphi, \varphi', \theta\sim\sqrt{\vep}$.
For the term involving Bessel functions, we need to make use of the Debye asymptotic behaviors
\begin{align*}
I_{\nu}(\nu z)\sim & \frac{1}{\sqrt{2\pi \nu}}\frac{e^{\nu\eta(z)}}{(1+z^2)^{\frac{1}{4}}}\left(1+\frac{u_1(t(z))}{\nu}+\ldots\right),\\
K_{\nu}(\nu z)\sim &\sqrt{\frac{\pi}{ 2 \nu}}\frac{e^{-\nu\eta(z)}}{(1+z^2)^{\frac{1}{4}}}\left(1-\frac{u_1(t(z))}{\nu}+\ldots\right),
\end{align*}
where
\begin{gather}
\eta(z)= \sqrt{1+z^2}+\log\frac{z}{1+\sqrt{1+z^2}},\hspace{1cm}
t(z)=\frac{1}{\sqrt{1+z^2}},\hspace{1cm}
u_1(t)=\frac{t}{8}-\frac{5t^3}{24},
\end{gather}and writing
\begin{align*}
\frac{I_{l+n_i}(\omega)- iI_{l+n_i+1}(\omega)}{K_{l+n_i+1}(\omega)+ iK_{l+n_i+1}(\omega)}
\sim & \frac{I_{l+n_i}(\omega)}{K_{l+n_i}(\omega)}\frac{\displaystyle 1-i\frac{I_{l+n_i+1}(\omega)}{I_{l+n_i}(\omega)}}
{\displaystyle1+i \frac{K_{l+n_i+1}(\omega)}{K_{l+n_i}(\omega)}}.
\end{align*}
After  integration, we  finally arrive at
\begin{align*}
E_{\text{Cas}}
\approx &-\frac{7\pi^3\hbar c R}{2880 d^2}\left(1+\left[\frac{1}{3}-\frac{20}{7\pi^2}\right]\frac{d}{R}\right).
\end{align*}

\section{VI. Proximity force approximation}
 The Casimir energy density on a pair of parallel plates  separated by a distance $d$ is given by 
 \begin{align*}
 \mathcal{E}_{\text{Cas}}^{\parallel}(d) =-\frac{7\pi^2\hbar c}{2880 d^3}.
 \end{align*}
 Hence, the proximity force approximation to the Casimir energy between a sphere and a plate is
\begin{align*}
E_{\text{Cas}}^{\text{PFA}}=&\iint_{x^2+y^2\leq R^2} dxdy \mathcal{E}_{\text{Cas}}^{\parallel}\left(L-\sqrt{R^2-x^2-y^2}\right)\\
=&2\pi \int_0^R dr\,r \mathcal{E}_{\text{Cas}}^{\parallel}\left(R+d-\sqrt{R^2-r^2}\right).
\end{align*}
Let $$v=\frac{R+d-\sqrt{R^2-r^2}}{d}.$$
Then
\begin{align*}
E_{\text{Cas}}^{\text{PFA}}=& 2\pi d\int_1^{(R+d)/d} dv\,(R+d-dv) \mathcal{E}_{\text{Cas}}^{\parallel}\left(dv\right)\\
\sim &2\pi R d\int_1^{\infty} dv \mathcal{E}_{\text{Cas}}^{\parallel}\left(dv\right)\\
\sim & -\frac{7\pi^3\hbar c R}{1440 d^2} \int_1^{\infty}dv\frac{1}{v^3}\\
=&-\frac{7\pi^3\hbar c R}{2880 d^2}.
\end{align*}
This coincides with the leading order term we obtain in the previous section.
\end{widetext}

\end{document}